\documentstyle[12pt]{article}
\def\simlt{\stackrel{<}{{}_\sim}}
\def\simgt{\stackrel{>}{{}_\sim}}
\topmargin
-2cm
\textwidth
15cm
\textheight
25cm
\oddsidemargin
1cm
\evensidemargin
1cm
\begin{document}
\pagestyle{empty}
\begin{titlepage}
\vspace{-4 in}
\rightline{CERN-TH/96-183}
\rightline{FTUAM 96/28}
\rightline{hep-ph/9607405}
\begin{center}
{\Large\bf Orbifold-induced $\mu$ term and
\\
\vskip .3 cm
electroweak symmetry breaking}
\end{center}
\vskip 1cm
\begin{center}
{\large A. Brignole$^1$,
 L.E. Ib\'a\~nez$^2$ and C. Mu\~noz$^2$}
\vskip 1cm
$^1$Theory Division, CERN  \\
CH-1211 Geneva 23, Switzerland
\vskip 0.8 cm
$^2$Departamento de F\'{\i}sica Te\'orica C-XI, \\
Universidad Aut\'onoma de Madrid, \\
Cantoblanco, 28049 Madrid, Spain
\\  
\end{center}
\vskip 2.5 cm
\begin{abstract}
\noindent
It is known that a Higgs $\mu$ term can be naturally 
generated through the K\"ahler potential in orbifold 
string models in which one of the 
three compactified complex planes has order two. 
In this class of models explicit expressions for both 
the $\mu$ parameter and the soft SUSY-breaking 
parameters can be obtained under the assumption that 
the goldstino is an arbitrary linear combination of 
the fermionic partners of the dilaton $S$ and all the 
moduli $T_i,U_i$. We apply this picture to the
MSSM and explore the consistency of the obtained
boundary conditions with radiative gauge symmetry 
breaking. We find that consistency with the measured 
value of the top-quark mass can only be achieved if
the goldstino has a negligible dilatino component
and relevant components along the $T_3,U_3$ moduli
associated to the order-two complex plane.
\end{abstract}
\vfill{
\noindent
CERN-TH/96-183
\newline
\noindent
July 1996}
\end{titlepage}
\newpage
\pagestyle{plain}
\setcounter{page}{1}
\vspace{5 mm}
{\bf 1.}
The Minimal Supersymmetric extension of the Standard Model
(MSSM) contains two kinds of mass terms: a set of soft 
supersymmetry-breaking terms, including scalar and
gaugino mass terms, and a globally supersymmetric
Higgs mass term, the so-called $\mu$ term.
If SUSY breaking originates from a super-Higgs mechanism in an 
underlying supergravity theory, the gravitino gets a mass $m_{3/2}$
and soft parameters ${\cal O}(m_{3/2})$ are usually generated.
In addition, if the supergravity K\"ahler potential contains 
appropriate terms bilinear in the Higgs fields, an effective 
$\mu$ parameter ${\cal O}(m_{3/2})$ can be generated as well 
\cite{GM}. Although not unique, this way to generate a $\mu$
term is particularly attractive because $\mu$ becomes directly 
related to SUSY breaking and thus stands essentially on the same 
footing as the other mass parameters.
 
In the restricted and motivated class of effective supergravity 
theories corresponding to 4-D superstring compactifications, 
specific patterns of soft terms emerge under the assumption that 
SUSY breaking is due to non-vanishing $F$-components for the dilaton ($S$)
and moduli ($T_i,U_i$) fields \cite{IL}--\cite{BIMS}. Such 
an assumption implies that the goldstino is a linear combination of the 
fermionic partners of the dilaton and the moduli, the coefficients of 
the combination just measuring the relative contribution of each field 
to SUSY breaking.  
We recall that, in the case of orbifold models, the set of K\"ahler moduli 
$T_i$ always includes the three diagonal K\"ahler moduli $T_1$,$T_2$,
$T_3$ associated to the three compactified complex planes, whereas the
number of complex structure moduli $U_i$ can be at most three. 
Here we will follow the approach of \cite{BIM,BIMS}, where the soft 
parameters are expressed in terms of the gravitino mass $m_{3/2}$, 
the angles specifying the (free) goldstino direction and the modular 
weights of the matter fields. However, we recall
that some ambiguity affects the results for the $\mu$ parameter 
and the associated soft $B$ parameter, depending on the source of 
the $\mu$ term itself. 
In this respect, an interesting and predictive class of models 
consists of orbifold models in which one of the three compactified 
complex planes (the third, say) has order two. We will focus on 
such a class of models and denote by $T_3$ and $U_3$ the K\"ahler and 
complex structure moduli associated to that plane. It was found in 
\cite{LLM,AGNT} that the K\"ahler potential corresponding to $T_3$ 
and $U_3$, and to charged untwisted fields $C_1$, $C_2$ in conjugate 
representations associated to the same plane, has the form
\begin{equation}
K =  - \log \left[ (T_3 + T_3^*) (U_3 + U_3^*) - 
(C_1 + C_2^*)(C_1^* + C_2) \right] \, .
\label{kk}
\end{equation}
After SUSY breaking, a certain effective $\mu$ term \cite{AGNT}
is induced for the fields $C_1$ and $C_2$, which we will identify here
with the electroweak Higgs fields. Under the assumption that this 
is the main source of the $\mu$ term and that SUSY breaking is 
dilaton/moduli-dominated, simple expressions for $\mu$ and 
$B$ in terms of goldstino angles can be derived \cite{BIMS}. 
Therefore a quite predictive scenario is obtained, with 
explicit and correlated expressions for the full set of 
mass parameters.
The aim of the present letter is to apply such a scenario 
to the MSSM and check its consistency with radiative 
electroweak symmetry breaking and the constraints on the 
top quark mass.

\vspace{5 mm}
{\bf 2.}
In the following we will assume that the MSSM can be obtained 
from a string model of the kind mentioned above and regard
the resulting tree-level expressions for the soft and $\mu$ 
parameters as boundary conditions given at some high scale $M_X$.
We will then use one-loop RGEs to evolve the parameters down to 
the electroweak scale $m_Z$, where we will impose the standard 
requirement of electroweak symmetry breaking and evaluate
the top mass. Such `leading-order' procedure, which neglects 
e.g. high- and low-energy threshold corrections, will be 
sufficient for our purpose. In the numerical evaluation, we 
will identify $M_X$ with the apparent unification scale 
of gauge couplings $M_G \sim 3 \times 10^{16}$ GeV, and 
check that the main conclusions do not change even if $M_X$
is allowed to take larger values.

We recall that the MSSM can be described in terms of a superpotential 
$W=\mu H_1 H_2 + h_t Q_3 U^c_3 H_2$ and a soft Lagrangian of the 
form
\begin{equation}
{\cal L}_{soft}= \left( \frac{1}{2} M_a \lambda^a \lambda^a - \mu B H_1 H_2
- h_t A_t Q_3 U^c_3 H_2 + {\rm h.c.} \right) 
- \sum_{\alpha} m_{\alpha}^2 |\phi_{\alpha}|^2  \, ,
\label{lsoft}
\end{equation}
where we have neglected Yukawa couplings different from the top 
one $h_t$.
The scalar potential, restricted to the real parts $h_1$, $h_2$ 
of the neutral Higgs fields, has the standard form:
\begin{equation}
 V(h_1,h_2) \ =\ (m_{H_1}^2+\mu^2) \, h_1^2 + (m_{H_2}^2+\mu^2) \, h_2^2
+(\mu B \, h_1 h_2+{\rm h.c.})\ + \frac{1}{8}(g^2+g'^2)(h_1^2-h_2^2)^2 \, ,
\label{pot}
\end{equation}
where $m_{H_1}^2$ and $m_{H_2}^2$ are the Higgs soft masses.
Under our assumptions the MSSM mass parameters are functions
of the goldstino direction, which can be parametrized\footnote{
We neglect complex phases and set the cosmological constant to 
zero.} by an angle $\theta$ and additional parameters $\Theta_i$ 
\cite{BIMS}. For instance, the goldstino components along the 
fermionic partners of the dilaton $S$ and the moduli 
$T_1$, $T_2$, $T_3$, $U_3$ are proportional to $\sin\theta$, 
$\cos\theta \, \Theta_1$, $\cos\theta \, \Theta_2$, 
$\cos\theta \, \Theta_3$, $\cos\theta \, \Theta'_3$, 
respectively. We also recall that
the parameters $\Theta_i$ are constrained by $\Theta_1^2 + 
\Theta_2^2 + \ldots + \Theta_3^2 + \Theta'^2_3=1$, where 
$\ldots$ correspond to possible contributions from
additional $T$-type or $U$-type moduli. However, it is
important to realize that the only goldstino parameters 
directly relevant for electroweak symmetry breaking are
$\theta$, $\Theta_3$ and $\Theta'_3$. First of all, such
parameters control the boundary conditions for the Higgs 
mass parameters. This is not surprising, of course, since 
the Higgs fields are associated to the third complex plane. 
Using the specific form (\ref{kk}) for the third-plane K\"ahler
potential, one finds that such mass parameters at the high scale 
$M_X$ can be expressed as \cite{BIMS}
\begin{eqnarray} 
& & m^2_{H_1}\, =\, m^2_{H_2}\, =\,  m_{3/2}^2\ \left( 1\ -\ 3\cos^2\theta
\, (\Theta_3^2+\Theta'^2_3)\right) 
\nonumber
\\
& &  \mu \, =\, m_{3/2}\ \left( 1\ +\ \sqrt{3}\cos\theta
\, (\Theta _3 + \Theta'_3)\right)  
\nonumber
\\
& & B \, = \, m_{3/2} \, 2 \, 
\frac{\left( 1+\sqrt{3} \cos\theta \, \Theta_3 \right)
\left( 1+\sqrt{3} \cos\theta \, \Theta'_3 \right)}
{\left( 1\ +\ \sqrt{3}\cos\theta \, (\Theta _3 + \Theta' _3)\right)} \, .
\label{softmx}
\end{eqnarray}
Second, the renormalization of the above parameters down to
$m_Z$ does not involve further goldstino angles. Indeed,
renormalization effects are controlled by $M_a$, $A_t$
and the specific combination $m_{\Sigma}^2 \equiv  m_{Q_3}^2 
+ m_{U_3}^2 + m_{H_2}^2$ of Higgs and stop soft masses.
The corresponding boundary conditions at $M_X$ are \cite{BIMS}
\begin{equation}
M_a \, = \,  - A_t \, = \, m_{3/2} \sqrt{3} \sin\theta  \,\, , \,\,
m_{\Sigma}^2 \, = \, m_{3/2}^2 \, 3 \, \sin^2\theta \, .
\label{auxmx}
\end{equation}
More comments on renormalization effects will be given below.
Here we recall for completeness that the above results for $A_t$ 
and $m_{\Sigma}^2$ (`sum rule') include the reasonable assumption 
that the coupling $h_t Q_3 U_3^c H_2$ corresponds to a renormalizable 
coupling also in the underlying string theory. Specifically, this 
implies that $Q_3$ and $U_3$ correspond to untwisted fields or twisted 
fields with overall modular weight $-1$ associated to the first and 
second complex planes. Notice, however, that we only need the combination 
$m_{\Sigma}^2$ and not the individual masses $m_{Q_3}^2$ and $m_{U_3}^2$,
which depend on the goldstino parameters $\Theta_1$, $\Theta_2$,\ldots, 
as well as on partial modular weights. We will not discuss the remaining 
squark and slepton masses either, since such a discussion would introduce 
further model dependence, i.e. a generic dependence on $\Theta_1$, 
$\Theta_2$,\ldots, and on individual modular weights \cite{BIMS}.
Such model-dependent features do not interfere with the issue of 
electroweak breaking, which is the main subject of the present analysis.

Before renormalizing the mass parameters, we summarize the situation
at the high scale $M_X$. We recall that the tree-level Higgs potential 
(\ref{pot}) at $M_X$ has a flat direction along $h_1=-h_2$, 
{\em independently of the goldstino direction}\footnote{An analogous 
result holds of course in other models, e.g. GUTs, having a conjugate 
pair of Higgs fields that could be identified with $C_1$ and $C_2$ 
in (\ref{kk}).} \cite{BIMS}. This property follows from the equality 
of the three coefficients
\begin{equation}
m_{H_1}^2+\mu^2 = m_{H_2}^2+\mu^2 = \mu B  \, ,
\label{equal}
\end{equation}
which control the quadratic part of the Higgs potential.
At this stage the  only constraint on the space of goldstino 
angles comes from the requirement that the Higgs potential be bounded 
along the orthogonal direction $h_1=h_2$, i.e. that the common value
of the three parameters in (\ref{equal}) be non-negative\footnote{Notice
that such a constraint does not prevent $m_{H_1}^2$ and $m_{H_2}^2$ from
being negative for some goldstino directions. For example, they are
indeed negative in the third-plane-dominated SUSY-breaking scenarios, 
which will turn out to be phenomenologically favoured.}:
\begin{equation}
\left( 1+\sqrt{3} \cos\theta \, \Theta_3 \right)
\left( 1+\sqrt{3} \cos\theta \, \Theta'_3 \right) \geq 0  \, .
\label{stab}
\end{equation}
Therefore, at the classical level, for {\em any} goldstino direction 
satisfying the stability condition (\ref{stab}) one finds
a continuous set of $SU(2)\times U(1)$ breaking vacua 
characterized by $\tan\beta=\langle h_2\rangle/
\langle h_1 \rangle =- 1$ and sliding Higgs vev's\footnote{A similar 
situation was found for specific goldstino directions and in a 
slightly different context in \cite{BFZ}.}. 
In order to remove such vacuum degeneracy we will adopt 
the conventional point of view that the electroweak vacuum 
is determined radiatively, i.e. by logarithmic and 
$h_t$-dependent quantum corrections which renormalize the parameters 
in the Higgs potential from $M_X$ down to $m_Z$. In so doing we will
encounter very different situations {\em depending on the 
goldstino direction}. 

The RG evolution of the parameters down to $m_Z$ can be 
performed numerically in a straightforward way. Analytical
formulae can be derived as well, e.g. along the lines 
of \cite{ILM}. Rather than giving the solutions in detail,
here we only note that the dependence of the renormalized 
parameters on the goldstino angles can be easily inferred
from the corresponding boundary conditions and the coupled 
system of RGEs. In particular, it is easy to see that $M_a$ and 
$A_t$ remain proportional to $\sin\theta$ and $m^2_{\Sigma}$ 
to $\sin^2 \theta$ at any scale. Therefore, since these 
parameters drive the RG evolution of $m_{H_1}^2$, $m_{H_2}^2$ 
and $B$, the additive renormalization of the latter parameters 
is controlled by $\sin\theta$ only\footnote{We are neglecting
terms proportional to $g'^2 \sum_{\alpha} Y_{\alpha} m_{\alpha}^2$ 
in the RGEs for scalar masses, which may induce a further small 
and model-dependent renormalization of $m_{H_1}^2$ 
and $m_{H_2}^2$, and can be considered
as part of our uncertainties. Notice however that such terms 
are absent when squark and slepton masses have universal boundary
conditions, e.g. in the
dilaton-dominated scenario and possibly others. Even in 
the general case, moreover, both $m_{\Sigma}^2$ and
$m_{H_1}^2+m_{H_2}^2$ are unaffected.}.
On the other hand, the $\mu$ parameter has a simple multiplicative 
renormalization as usual. In conclusion, the Higgs potential mass 
parameters at the scale $m_Z$ have the form
\begin{eqnarray}
 m^2_{H_1} & = &  m_{3/2}^2\ \left( 1\ -\ 3\cos^2\theta
\, (\Theta_3^2+\Theta'^2_3) + \sin^2\theta \, c_1 \right) 
\nonumber
\\
 m^2_{H_2} & = &  m_{3/2}^2\ \left( 1\ -\ 3\cos^2\theta
\, (\Theta_3^2+\Theta'^2_3) + \sin^2\theta \, c_2 \right) 
\nonumber
\\
 \mu \ &  = & \ m_{3/2}\ \left( 1\ +\ \sqrt{3}\cos\theta
\, (\Theta _3 + \Theta' _3)\right) c_{\mu}
\nonumber
\\
B & = & m_{3/2} \left[ 2 \, \frac{\left( 1+\sqrt{3} \cos\theta 
\, \Theta_3 \right) \left( 1+\sqrt{3} \cos\theta \, \Theta'_3 \right)}
{\left( 1\ +\ \sqrt{3}\cos\theta \, (\Theta _3 + \Theta' _3)\right)}
+ \sin\theta \, c_B \right]
\label{softmz}
\end{eqnarray}
where $c_i$ are calculable renormalization coefficients\footnote{
Alternatively to a numerical evaluation, one could use e.g. the 
analytical formulae in the second ref. of \cite{ILM} and express 
the $c_i$'s as 
$c_1=3g$, $c_2=h+3(e+f-k)-1$, $c_{\mu}=q$, $c_B=\sqrt{3}(r+s)/q$,
where the opposite sign convention for gaugino masses has
been taken into account. Note that our sign conventions in
(\ref{lsoft}) lead to RGEs for $A$ and $B$ having signs as
in $dA/d t \sim -A - M$, $d B/ d t  \sim - A - M$,
where $t = \log M_X^2/Q^2$.}
which depend on $\log(M_X/m_Z)$ and (apart from $c_1$)
on the boundary value for $h_t$, which we denote by $h_t^0$. 
The $c_i$'s do not depend on the goldstino angles, which are 
explicitly factored out in (\ref{softmz}).

For a given goldstino direction satisfying (\ref{stab}), 
radiative electroweak symmetry breaking occurs if the Higgs
potential mass parameters at $m_Z$ satisfy the basic conditions
\begin{eqnarray}
& & (m_{H_1}^2 + \mu^2 ) (m_{H_2}^2 + \mu^2 )  < (\mu B)^2
\nonumber
\\
& & (m_{H_1}^2 + \mu^2 )+(m_{H_2}^2 + \mu^2 )  \geq 2 |\mu B|
\label{sbc}
\end{eqnarray}
which express the instability of the Higgs potential 
at the origin and its boundedness along the D-flat directions, 
respectively. Since $m_{3/2}$ is only an overall factor
in (\ref{sbc}), the relevant parameter to be adjusted is just 
$h_t^0$. Typically, the conditions (\ref{sbc}) are simultaneously 
satisfied only for a narrow range of $h_t^0$ values, which is mapped
onto an even narrower range of $h_t(m_Z)$ values. For $h_t^0$
in the favourable range the Higgs potential develops an 
$SU(2)\times U(1)$ breaking minimum. The corresponding
$\tan\beta$ can be computed from\footnote{As a side remark,
we note that in the present approach the signs of all input 
parameters are given and therefore the output parameter 
$\tan\beta$ will have a sign too. The correct sign can be 
obtained using e.g. $\tan\beta = (1-\cos 2\beta)/\sin 2\beta$, where
$\cos 2\beta = - \sqrt{1-\sin^2 2\beta}$ as a consequence
of $m_{H_2}^2 < m_{H_1}^2$.} 
\begin{equation}
\sin 2\beta = \frac{- 2 \mu B}{m_{H_1}^2 + m_{H_2}^2 + 2 \mu^2 } \, ,
\label{sin}
\end{equation}
where $m_{3/2}$ drops out again. Finally one can evaluate
the corresponding top mass parameter ${\hat m_t} = v h_t(m_Z) 
|\sin\beta|$, with $v \simeq 174$ GeV. In conclusion, the
symmetry breaking requirement allows us to associate a certain range 
of ${\hat m_t}$ values to any given goldstino direction.  
Rather than translating ${\hat m_t}$ values into pole mass
values, we will be content with a qualitative comparison
with the experimental constraints and consider ${\hat m_t}$ 
values smaller than 150 GeV as phenomenologically 
unacceptable. We will see that this simple and conservative 
criterion will be sufficient to put strong constraints on the
different SUSY-breaking scenarios. 

As an additional comment, we note that we have exploited
only one of the two relations that follow from the
minimization of the Higgs potential, i.e. eq.~(\ref{sin}) above.
The second relation can be written as
\begin{equation}
m_Z^2 = \frac{m_{H_2}^2 - m_{H_1}^2}{\cos 2\beta} -
(m_{H_1}^2 + m_{H_2}^2 + 2 \mu^2)
\label{cos}
\end{equation}
with $\beta$ given in (\ref{sin}). Since $m_{3/2}^2$ is an overall 
factor in the right-hand side, the above equation formally 
establishes a one-to-one correspondence between $m_Z^2/m_{3/2}^2$
and $h_t^0$ for any given goldstino direction. Therefore,
it seems that the minimization conditions in principle allow 
the scale $m_{3/2}$ to be fixed as a function of $h_t^0$ (or 
${\hat m_t}$). However, such correspondence is of little practical
use, mainly because a narrow range of allowed $h_t^0$
values corresponds to a wide (formally infinite) range
of  $m_{3/2}$ values and the mapping itself is subject
to several uncertainties. Therefore we will not rely
explicitly on such $m_{3/2}$--${\hat m_t}$ correlation.

\vspace{5 mm}
{\bf 3.}
For convenience, we will describe our results in terms
of three categories of goldstino directions, each one
being a special case of the next: i) pure dilaton 
SUSY breaking; ii) dilaton/overall-modulus SUSY breaking; 
iii) dilaton/moduli SUSY breaking.

i) {\em Pure dilaton SUSY breaking.} This scenario corresponds 
to the limit $\cos\theta \rightarrow 0$, i.e. to the 
two inequivalent cases $\sin\theta = \pm 1$. 
The stability condition (\ref{stab}) is automatically
satisfied. The renormalized mass parameters (\ref{softmz})
read
\begin{eqnarray}
m^2_{H_1} & = &  m_{3/2}^2 \, (1 + c_1) 
\nonumber
\\
m^2_{H_2} & = &  m_{3/2}^2 \, (1 + c_2)
\nonumber
\\
\mu &  = &  m_{3/2} \, c_{\mu}
\nonumber
\\
B & = & m_{3/2} \, ( 2 \pm c_B ) \, .
\end{eqnarray}

The numerical analysis shows that for $\sin\theta = + 1$
the symmetry-breaking requirements 
(\ref{sbc}) select the initial top Yukawa in the range 
$h_t^0 \sim $ (0.11--0.13). Correspondingly, one finds
$h_t (m_Z) \sim $ (0.4--0.5), $\tan\beta \sim - $(1--3)
and too small a top mass ${\hat m}_t < 70$ GeV. 
For $\sin\theta = - 1$, the situation is even worse.
Proper symmetry breaking occurs for $h_t^0 \sim$ (0.04--0.05),
corresponding to $h_t(m_Z) \sim 0.2$, $\tan\beta \sim - $(1--1.2)
and ${\hat m}_t < 25$ GeV. Therefore the specific $\mu$ mechanism
considered here is incompatible with the dilaton SUSY-breaking
scenario. This result is consistent e.g. with the findings of ref.
\cite{BLM}, where the dilaton scenario phenomenology was studied using 
the boundary condition $B = 2 m_{3/2}$ whereas $\mu$ was treated 
as a free parameter. In such an approach the value of $\mu$ 
{\em required} by radiative symmetry breaking and leading to 
an acceptable top mass was much larger than the specific 
`orbifold' value $\mu = m_{3/2}$ used here.

ii) {\em Dilaton/overall-modulus SUSY breaking.} This
scenario corresponds to generic $\theta$ and to
$\Theta_1=\Theta_2=\Theta_3=1/\sqrt{3}$ (implying in 
particular $\Theta'_3=0$), i.e. to the limit 
in which only the dilaton and a symmetric combination 
of the K\"ahler moduli $T_1$, $T_2$, $T_3$ participate
to SUSY breaking. Again, the stability condition 
(\ref{stab}) is automatically satisfied. The renormalized 
mass parameters (\ref{softmz}) now read
\begin{eqnarray}
m^2_{H_1} & = &  m_{3/2}^2 \, \sin^2\theta \, (1 + c_1) 
\nonumber
\\
m^2_{H_2} & = &  m_{3/2}^2 \, \sin^2\theta \, (1 + c_2)
\nonumber
\\
\mu &  = &  m_{3/2} \, (1+\cos\theta) \, c_{\mu}
\nonumber
\\
B & = & m_{3/2} \, ( 2 + \sin\theta \, c_B )
\end{eqnarray}
and the symmetry-breaking requirements (\ref{sbc}) 
associate a certain range of $h_t^0$ values to each $\theta$.
Such $h_t^0$ values are quite small in most of the $\theta$ range, 
typically implying a top mass below 100 GeV. For 
$-30^\circ \simlt \theta \simlt 30^\circ$ one can get somewhat
higher values for $h_t^0$ ($\sim$ 0.3--0.6), leading 
to $h_t(m_Z)$ values close to 1. In such cases, however,
$|\tan\beta|$ turns out to be very close to 1 and 
one always finds ${\hat m}_t \simlt$~140 GeV, the
maximum value being reached for $\theta \sim - 20^\circ$.
Therefore also this class of goldstino directions
seems incompatible with radiative breaking and
top mass value. Notice, however, that the situation
improves when one moves from dilaton- to moduli-
dominated SUSY breaking. 
  
iii) {\em Dilaton/moduli SUSY breaking.} This is the
most general scenario, corresponding to generic 
values for $\theta$ and $\Theta_i$, i.e. a generic 
goldstino direction. Actually we expect the potentially
interesting directions to have small $\sin\theta$, since
the analysis of scenario (ii) has shown that an improvement
can be obtained only when the goldstino has a suppressed
dilatino component.
To obtain a further improvement one should depart from
the overall-modulus direction and explore different
ways to distribute the non-dilatonic SUSY-breaking contribution 
among the individual moduli. In this respect, we could
expect $T_3$ and $U_3$ to give a peculiar 
contribution to SUSY breaking, since the third complex 
plane plays a special role in the class of models
considered here. For instance, one can think of the two 
limiting and complementary situations in which those moduli 
dominate SUSY breaking or do not participate at all.
These will be special and important examples of the
general scenario considered here. Before reporting the general
results, we note that the goldtsino directions are now subject to 
two preliminary constraints. One is the stability
condition (\ref{stab}), which is no longer automatic.
An additional constraint arises from the sum rule relating 
the boundary values of stop and Higgs soft masses.
If one requires $m_{Q_3}^2 + m_{U_3}^2 \geq 0$ at $M_X$ in order 
to avoid instabilities of the scalar potential along
charged and coloured directions, the formulae for
$m_{\Sigma}^2 = m_{Q_3}^2 + m_{U_3}^2 + m_{H_2}^2$ and $ m_{H_2}^2$ 
in (\ref{auxmx}) and (\ref{softmx}) lead to the constraint
\begin{equation}
\Theta_3^2 + \Theta'^2_3 \geq 1 - \frac{2}{3 \cos^2\theta} \, .
\label{qu}
\end{equation}
This inequality was automatically satisfied for the
special goldstino directions considered in (i) and (ii).
Here it constrains the parameters $\Theta_3$, $\Theta'_3$
when $\cos^2\theta > 2/3$, or equivalently the angle $\theta$ 
when $\Theta_3^2 + \Theta'^2_3 < 1/3$. We recall that 
$\Theta_3^2 + \Theta'^2_3 \leq 1$ by definition. 

The general form of the renormalized Higgs potential mass 
parameters has already been written above (\ref{softmz}). 
Once the radiative symmetry-breaking
conditions are imposed, we find that the top mass ${\hat m}_t$
is too small in the case of negligible $\Theta_3$, $\Theta'_3$,
i.e. when the third plane moduli do not participate significantly
to SUSY breaking. This result is partly due to the above
constraint (\ref{qu}), which prevents us from reaching the small 
$\sin\theta$ region\footnote{If that constraint is relaxed, some points
with large ${\hat m}_t$ values can be found, but we will
not take this possibility into account in the present discussion.}.
One can get ${\hat m}_t \simgt 150$ GeV only
if $\sin\theta$ is sufficiently small and at the same time
{\em both} $\Theta_3$ and $\Theta'_3$ are  non-negligible.
Roughly speaking, the former condition allows us to reach sufficiently
large values for $h_t^0$ (anyway smaller than $\sim 0.6$,
but this is enough), whereas the latter one allows us to
displace $|\tan\beta|$ from 1 at the same time. 
For a heuristic explanation of the latter point, we recall 
that $|\tan\beta|\gg 1$ corresponds to $|\sin 2\beta|\ll 1$, 
which typically requires $|B| \ll m_{3/2}$. For small $\sin\theta$ 
this cannot happen in the overall-modulus case ($\Theta_3=1/\sqrt{3}$,
$\Theta'_3=0$) because $B$ is dominated by the boundary condition 
$B = 2 m_{3/2}$, whereas it can happen for simultaneously 
non-vanishing $\Theta_3$, $\Theta'_3$. As an example
of favourable goldstino direction, we can quote e.g. the 
limiting case $\Theta_3=\Theta'_3=1/\sqrt{2}$, where $T_3$ and $U_3$ 
are the only moduli contributing to SUSY breaking\footnote{It is 
interesting to apply the radiative breaking criterion to the explicit 
simple model built in the first ref. of \cite{BFZ}, which includes
a K\"ahler potential of the form (\ref{kk}). 
There the only modulus contributing to SUSY breaking is the diagonal 
combination of $T_3$ and $U_3$, corresponding just to the favourable
values $\Theta_3=\Theta'_3=1/\sqrt{2}$. However in such a model the 
dilaton SUSY-breaking contribution is non-negligible. It corresponds
to $\sin \theta=-1/\sqrt{3}$ ($\cos\theta = -\sqrt{2/3}$), which leads 
to too small a top mass, ${\hat m}_t <$ 80 GeV.}. 
Large values of ${\hat m}_t \sim$ 150--175 GeV can
be reached for $\theta \sim $ ($170^\circ$--$180^\circ$), 
the maximum ${\hat m}_t$ corresponding to $\theta \sim 176^\circ$. 
A similar situation can be found in a nearby region and in another
not equivalent region where sign$(\Theta_3 \Theta'_3)<0$. For instance,
the latter region does not include the values 
$\Theta_3=-\Theta'_3=1/\sqrt{2}$ because the constraint (\ref{stab}) 
is violated for small $\sin\theta$.
Also, there the largest ${\hat m}_t$ values are obtained for 
$\cos\theta >0$ instead of $\cos\theta<0$.
However we refrain from giving a detailed correspondence between 
goldstino angles $\{ \theta, \Theta_3, \Theta'_3 \}$
and values of ${\hat m}_t$, since such mapping
is affected by several uncertainties\footnote{We recall that,
among other effects, we have neglected string loop 
corrections. These may become relevant precisely
in the small $\sin\theta$ region, but they are
very model-dependent.}. We rather stress again the main 
conclusion of the exploration in the general case, that is
the possibility to reach phenomenologically acceptable values 
of the top mass in a region of goldstino angles characterized 
by small $\sin\theta$ and simultaneously non-negligible 
$\Theta_3$, $\Theta'_3$. 

We conclude this section by mentioning some qualitative features
of the SUSY spectra corresponding to the favourable 
region of goldstino angles. In such a region the boundary 
conditions (\ref{auxmx}) for gaugino masses and $A_t$ are 
smaller than $m_{3/2}$, due to the smallness of $\sin\theta$, 
whereas the boundary conditions for stop masses and $\mu$
are generically ${\cal O}(m_{3/2})$. 
This hierarchy tends to be preserved at low energy. 
In particular, the lightest chargino and neutralinos are
mainly gaugino-like and typically lighter than the stop squarks,
whereas the heaviest ones are mainly higgsino-like and 
have masses of the same order as the stop squarks. 
The gluino and the heavy Higgses tend to stay in an 
intermediate range. We stress again that the precise mass 
ratios will depend on the specific goldstino direction.
We also recall that general predictions for the remaining 
squark and slepton masses cannot be made unless
additional model-dependent information is provided,
as mentioned in section 2. 

\vspace{5 mm}
{\bf 4.}
In summary, we have applied the general results
of the dilaton/moduli SUSY-breaking scenario 
to the MSSM, using at the same time a specific and
predictive mechanism for generating $\mu$ available in
a class of orbifold models. Such models have an order-two
compactified complex plane and induce an effective $\mu$
parameter through Higgs bilinear terms in the K\"ahler
potential. Starting from such `stringy' boundary conditions 
for the MSSM mass parameters, we have studied 
whether the combined requirement of radiative 
electroweak symmetry breaking and a sufficiently
large top mass constrain the goldstino direction,
i.e. the relative size of the dilaton and moduli 
$F$-terms. We have found that the above phenomenological
requirements cannot be satisfied either in the dilaton
dominated scenario or in the mixed dilaton/overall-modulus 
scenario. The only SUSY-breaking scenario 
compatible with such constraints requires a suppressed
dilaton contribution and important (often dominant)
contributions from the moduli fields associated
to the same order-two plane as the Higgs fields. On the phenomenological
side such a restricted SUSY-breaking scenario implies certain
features for the SUSY spectra. On the theoretical side 
the above conclusion can perhaps be taken as an indication 
guiding the search for explicit models of SUSY breaking.


\begin{thebibliography}{99}
%
\bibitem{GM} G.F. Giudice and A. Masiero, {\it Phys. Lett.} {\bf B206}
(1988) 480.
%
\bibitem{IL} L.E. Ib\'{a}\~{n}ez and D. L\"ust,
{\it Nucl. Phys.} {\bf B382} (1992) 305.
%
\bibitem{KL} V.S. Kaplunovsky and J. Louis,
{\it Phys. Lett.} {\bf B306} (1993) 269.
%
\bibitem{BIM} A. Brignole, L.E. Ib\'{a}\~{n}ez and C. Mu\~noz,
{\it Nucl. Phys.} {\bf B422} (1994) 125 [Erratum: {\bf B436} (1995) 747].
%
\bibitem{FKZ} S. Ferrara, C. Kounnas and F. Zwirner,
{\it Nucl. Phys.} {\bf B429} (1994) 589 [Erratum: {\bf B433} (1995) 255].
%
\bibitem{otros} T. Kobayashi, D. Suematsu, K. Yamada and Y. Yamagishi,
{\it Phys. Lett.} {\bf B348} (1995) 402;
\\
P. Brax and M. Chemtob, {\it Phys. Rev.} {\bf D51} (1995) 6550;
\\
E. Dudas, S. Pokorski and C.A. Savoy, {\it Phys. Lett.} {\bf B369}
(1996) 255.
%
\bibitem{BIMS} A. Brignole, L.E. Ib\'{a}\~{n}ez, C. Mu\~noz 
and C. Scheich, hep-ph/9508258, to appear in {\it Z. Phys. C.}
%
\bibitem{LLM} G. Lopes-Cardoso, D. L\"ust and T. Mohaupt,
{\it Nucl. Phys.} {\bf B432} (1994) 68.
%
\bibitem{AGNT} I. Antoniadis, E. Gava, K.S. Narain and T.R. Taylor,
{\it Nucl. Phys.} {\bf B432} (1994) 187.
%
\bibitem{BFZ} A. Brignole and F. Zwirner, {\it Phys. Lett.} {\bf B342}
(1995) 117;
\\
A. Brignole, F. Feruglio and F. Zwirner,
{\it Phys. Lett.} {\bf B356} (1995) 500.
%
\bibitem{ILM}
L.E. Ib\'{a}\~{n}ez and C. Lopez, {\it Nucl. Phys.} {\bf B233} (1984) 511;
\\
L.E. Ib\'{a}\~{n}ez, C. Lopez and C. Mu\~noz, {\it Nucl. Phys.} 
{\bf B256} (1985) 218.
%
\bibitem{BLM} R. Barbieri, J. Louis and M. Moretti,
{\it Phys. Lett.} {\bf B312} (1993) 451.
%
\end{thebibliography}
\end{document}